\documentclass[12pt]{article}
\input epsf
\begin{document}

\begin{flushright}
RU--96-43\\
\today\\
\end{flushright}

\begin{center}

\vspace{0.5cm}
{\Large \bf Pseudoscalar Glueball Mass:\\ 
QCD vs. Lattice Gauge Theory Prediction}

\vspace{0.5in}

{\bf Gregory  Gabadadze }

\vspace{0.2in}

{\baselineskip=14pt
Department of Physics and Astronomy, Rutgers University,\\
Piscataway, New Jersey 08855-0849, USA}

{email: gabad@physics.rutgers.edu}
\vspace{1.5cm}

{\bf Abstract}
\end{center}
We study whether the pseudoscalar glueball mass in full QCD can differ
from the prediction of quenched lattice calculations. 
Using properties of the correlator
of the vacuum topological susceptibility we derive an expression 
for the upper bound on  the QCD glueball mass.
We show that the QCD pseudoscalar glueball is lighter than
the pure Yang-Mills theory glueball studied in 
quenched lattice calculations.    
The mass difference between those two states
is of order of $1/N_c$. The value calculated for the $0^{-+}$ QCD 
glueball  mass can not be reconciled  with 
any physical state observed so far in the corresponding channel. 
The glueball decay constant and its production rate in  $J/\psi$ radiative
decays  are calculated. The production rate is large enough to be 
studied experimentally.

\vspace{1.0in}


\newpage
\begin{center}
{\bf Introduction} 
\end{center}

Glueballs are one of the intriguing theoretical predictions 
of QCD \cite{Gell}. 
The search for these composites is a long-standing problem of 
theory and experiment.
We now have  detailed 
experimental studies  of the resonances  
in the mass region  up to  $2.3$ GeV, as well as  
great progress in lattice QCD calculations. 
A number of
interesting particles have been detected \cite{PDG}. Some of them,
the $\eta (1410, 0^{-+})$, the $f_0(1500, 0^{++})$,
the $f_0(1710, 0^{++})$ and the $\zeta  (2230, 2^{++})$
appear to have  a rich gluon content.
(For a recent discussion of the phenomenology of these 
composites and  a full set of  references see   \cite{CFL}).  
The glueball candidates can be compared to  
lattice QCD predictions\footnote{
The earliest theoretical predictions were based on  QCD
sum rule \cite{SVZ} calculations \cite{NSVZ1}-\cite{Grunberg}.}.
These calculations argue in favor of the 
following hierarchy of glueball masses:  the $0^{++}$ glueball 
is the lightest one 
with  mass  about $1.5-1.7$
GeV \cite{UKQCD}, \cite{Wein};  the $2^{++}$ state, 
having mass  $2-2.2$ GeV \cite{UKQCD}, is the next one; finally, 
the $0^{-+}$ pseudoscalar glueball, being predicted 
in the lattice calculations to have mass  $2.3\pm 0.2$ GeV \cite{UKQCD}
is the heaviest one. 
Various studies of pure Yang-Mills (YM) theory also support 
the picture outlined above. In ref. \cite{West} the theorem was proved
that in  pure gluodynamics the pseudoscalar glueball
is heavier than the scalar one. Instanton
calculations \cite{Shuryak} also  confirm this picture.

Notice that all the theoretical facts listed above  are 
firmly established 
results of pure YM theory. One might wonder whether this picture 
is affected  when quarks are also included in the theory.
Recent analyses \cite {Wein}, \cite {AmslerClose}, \cite {CFL}, 
\cite {AAS} of the  scalar glueball  
candidates indicate  an important 
mixing with  the nearby ${\bar q} q$ resonances. 
This leads to  a modification of the  mass spectrum and the 
decay constants  of the glueball states \cite {CFL}, \cite {AAS}. 
The mass shifts due to mixing are 
approximately $100~{\rm MeV}$ or so, 
and the lattice predictions are in good agreement at this level 
with the experimental data for the $0^{++}$ and the $2^{++}$ channels
\cite {CFL}, \cite {AAS}. 

However, the situation  
for the pseudoscalar channel is problematic. 
One can expect that, because of the axial anomaly,  quarks  
are  crucial for $0^{-+}$ channel physics \cite{'t Hooft}
As we mentioned above,
the lightest $0^{-+}$ glueball predicted by the lattice calculations
has mass  about $2.3$ GeV \cite{UKQCD}. On the other hand,
there is evidence  that the $\eta(1410)$ can have a substantial 
gluonic component  \cite {CFL}
\footnote{Note that structure originally identified as 
$\eta(1440)$  is in fact two states, the $\eta(1410)$ and the $\eta(1490)$,
\cite {MarkIII48}-\cite
{CBC16}. The lighter $\eta(1410)$ seems to have a rich gluon content
\cite {CFL}, while
the heavier $\eta(1490)$ is dominantly ${\bar s } s$ state
possibly with some admixture of glue \cite{CFL}.}.
Hence,  an important  question is how the 
quark degrees of freedom may shift the glueball 
mass in the $0^{-+}$ channel and  whether one can identify
the QCD glueball  with any observed state. 
We examine here these questions. 

In low energy QCD  the
$\eta'(958)$ makes  the  dominant contribution to  correlators
in the $0^{-+}$ flavor singlet channel. It is well known that the 
mass and  decays of the  $\eta'$ meson are  strongly affected by the 
gluonic sector of the theory.  
The axial anomaly and  instantons  
play a  key role in generating the mass and  decay constant
of the $\eta'$ \cite{'t Hooft} (see also the papers 
\cite{NSVZ1}, \cite{BartTat}, \cite{BallFT} and refs. therein). 
This suggests that the pure $0^{-+}$ glueball 
should be also affected  by 
the singlet ${\bar q} q$
admixture in full QCD \cite{FN}. 
In what follows we will refer to this mixed 
glueball-${\bar q}q$ state as a  QCD glueball in distinction
with the pure YM glueball and the physical $\eta'$.

We are going to  derive an inequality  between the mass 
of the QCD pseudoscalar glueball 
and the mass of the YM glueball 
as measured in quenched lattice calculations \cite {UKQCD}.
In the  derivation we closely follow  the  argumentation
of  Witten \cite{Witten} and Veneziano \cite{Venezia}, but 
keep truck of the finite $N_c$ effects by using the method of  QCD sum rules 
\cite {SVZ}. The decay constant and mass for the QCD 
pseudoscalar glueball will be determined. We show that 
the QCD glueball is lighter then the one of pure  YM theory. 
Finally, we will look  at  $J/\psi$ radiative decays 
and predict the production rate for the QCD glueball state.
\begin{center}
{\bf 1. QCD Sum Rules and the Witten-Veneziano Relation}
\end{center}

In this section  we are going to study the properties of  the 
correlator of the vacuum topological
susceptibility. In Minkowski space-time this  is defined as 
\begin{eqnarray}
\chi(q^2\equiv -Q^2)=\left ({g^2\over 32 \pi^2} \right )^2i \int e^{iqx}
\langle 0|T G^a_{\mu\nu}{\tilde G^a_{\mu\nu}}(x) G^b_{\alpha\beta}{\tilde
G^b_{\alpha\beta}}(0)|0 \rangle d^4x,        
\end{eqnarray}
where ${\tilde G^a_{\mu\nu}}={1\over 2}\varepsilon_{\mu\nu\alpha\beta}
G^a_{\alpha\beta}$ and ${g^2\over 32 \pi^2}G_{\mu\nu}{\tilde
G_{\mu\nu}}$ is the renormalized composite operator with 
$g$ being the running coupling constant. 

This correlator has a different behavior 
in the vicinity of  $Q^2=0$
depending on whether it is evaluated  in full  QCD or  in pure 
Yang-Mills theory \cite{Witten}. If light quarks are 
included in the theory,
$\chi_{\rm QCD}(0)$ is proportional to the  product of the quark masses
and  vanishes in the chiral limit. This fact
is related to the absence of the theta angle in massless QCD. 
In general, the theta  angle can always be rotated away by an 
appropriate chiral transformation of a massless fermionic field. 
On the other hand, there are no massless fermions 
in pure YM theory. As a result,  the   explicit theta 
dependence cannot be removed \cite{Witten} and 
$\chi_{\rm YM}(0)$ turns out to be a  nonzero number. 

Let us consider the dispersion relation for the function $\chi(Q^2)/Q^2$
\begin{eqnarray}
{\chi(0)-\chi(Q^2)\over Q^2}={1\over \pi}\int_0^{\infty}{\rho(s) ds
\over s(s+Q^2)}+{\rm subtractions}. 
\end{eqnarray} 
Following the standard 
QCD sum rule approach \cite{SVZ}, the spectral density for this
correlator, $\rho$, can be
decomposed into two parts. The first one consists of resonance (pole)
contributions and the second one is determined by the perturbative 
expansion
\begin{eqnarray}
\rho(s)=\rho^{poles}(s)+{\tilde \rho}(s/\mu^2)\theta(s-s_0(\mu)),~~~~~
\rho^{poles}(s)\equiv \sum_n c_n\delta (s-m_n^2),
\nonumber
\end{eqnarray} 
where the $c_n$'s are resonance residues, $m_n$'s are corresponding
masses,  and $s_0$ denotes  the continuum threshold; 
${\tilde \rho}$ is given by 
the perturbative expansion of the corresponding correlator.
We are going to work in  the next-to-leading order (NLO)
of the perturbative expansion.
In the case at hand, the leading contribution to  ${\tilde \rho}$
is scale and scheme independent. However,  the next-to-leading
term depends on the renormalization scheme.
In leading  order  
${\tilde \rho}$ is fixed  by the diagram which contains
only gluon propagators. This diagram is the same  for  
QCD and pure Yang-Mills theory. 
Hence, in lowest order 
$ {\tilde \rho}_{\rm QCD}={\tilde \rho}_{\rm YM}.$ 
However, this relation does not hold
in the NLO. There are quark loop corrections to the 
gluon propagator in QCD.  Hence,  the result for $ {\tilde \rho}$  
in NLO QCD differs  from the one of pure YM theory  
by quark loop contributions. 
In order to fix the scale/scheme ambiguity of the NLO corrections it is 
convenient  to adopt  the Brodsky-Lepage-Mackenzie (BLM) scale fixing
procedure \cite{BLM}.  
In the BLM scheme NLO quark loop insertions into the gluon propagator
are summed up into the redefinition of the effective scale of the  strong
coupling constant. Hence, in the BLM scheme, the perturbative
expansion for $ {\tilde \rho}_{\rm QCD}$  in the next-to-leading order 
formally coincides
with that  for  $ {\tilde \rho}_{\rm YM}$.  
Keeping in mind that we have adopted the BLM scheme 
one can write down the following relation 
for the spectral densities in QCD and pure Yang-Mills theory
\footnote{In this  case the BLM scheme
leads to the better convergence of the perturbation 
expansion.   
The  NLO corrections are 
large in the ${\overline {MS}}$ 
scheme \cite {KKP}.} \cite {KKP}
\begin{eqnarray}
{\tilde \rho}_{\rm QCD}={\tilde
\rho}_{\rm YM}
=\left ({g^2\over 32
\pi^2}\right )^2 
{2 \over \pi}(1+5 {\alpha_s\over \pi})s^2\equiv as^2.
\end{eqnarray}
Before we turn to the resonance part of the spectral density 
let us make an important comment. It deals with the
definition of the running coupling constant in QCD and in YM theory. 
The expression for $\alpha_s(s/\mu^2)$
depends on the number of flavors $N_f$ present in the theory. 
In pure YM theory $N_f=0$ and the coupling constant 
of this theory differs from the one defined in full QCD.  
However, our goal is to stay maximally
close to what is used in quenched lattice calculations. 
In those calculations only gluon degrees of freedom are taken into
account. But this is not the whole story.
Quenched lattice calculations 
effectively include some of the virtual 
quark effects  through the formal substitution 
of the QCD running coupling with $N_f=3$ instead of the coupling  
of YM theory with $N_f=0$  (see ref. \cite {Wein} 
for this discussion). 
We are using this formal method through the 
paper. In particular, 
the theory to which we refer as pure YM is actually the theory  
with some of the virtual quark loops  effectively included through the 
use of the QCD running coupling constant $\alpha_s$ instead of the 
pure YM running coupling. 
Thus, YM theory in our context refers to the model which has 
the full QCD coupling constant, but nevertheless,
differs from  true QCD by the absence of ${\bar q} q$ 
bound states and  the absence of quark condensate effects. 
Let us stress again that
these conventions differ form the normally used ones (adopted  
for example in ref. \cite {NSVZbig})
and are motivated by the quenched lattice calculation procedure.

After this remark let us turn to the 
resonance part of the spectral density.
This part for YM theory is assumed to be 
saturated by 
the pure glueball state $G_0$,  and for QCD by the $\eta'$ meson 
and  QCD glueball state $G$. The expressions for the spectral densities 
are    
\begin{eqnarray}
\rho_{\rm YM}(s)=f^2_{G_0}m^4_{G_0}\delta(s-m^2_{G_0})+
{\tilde \rho}_{\rm YM}(s)\theta(s-s_{0}), \nonumber \\
\rho_{\rm QCD}(s)=f^2_{G}m^4_{G}\delta(s-m^2_{G})+
f^2_{\eta '}m^4_{\eta'}\delta(s-m^2_{\eta'}) 
+{\tilde\rho}_{\rm QCD}(s)\theta(s-s_{1}),
\end{eqnarray}
where $s_0$ and $s_1$ denote the continuum thresholds
for YM theory and full QCD respectively. 
Other quantities in these equations are defined as follows 
\begin{eqnarray}
\langle 0|{g^2\over 32 \pi^2} G^a_{\mu\nu}{\tilde G^a_{\mu\nu}}|G_0\rangle
=f_{G_0}m_{G_0}^2,~~~
\langle 0|{g^2\over 32 \pi^2} G^a_{\mu\nu}{\tilde G^a_{\mu\nu}}|G\rangle
=f_{G}m_{G}^2, \nonumber \\
\langle 0|{g^2\over 32 \pi^2} G^a_{\mu\nu}
{\tilde G^a_{\mu\nu}}|\eta'\rangle
=f_{\eta'}m_{\eta'}^2,~~~~~~~~~~~~~~~~~~~~~~
\end{eqnarray}
with $m_G$ being the QCD glueball mass, $m_{G_0}$  being the
YM glueball mass, and $f_{G}$ and $f_{G_0}$  the corresponding
decay constants\footnote{The gluonic operator 
${g^2\over 32 \pi^2}G{\tilde G}$  appearing in eq. (5) has an
anomalous dimension  so that the constants 
$f_{G_0}$, $f_{G}$ and  $f_{\eta'}$  
are not renormalization 
group invariant quantities \cite {SV}. 
For any of three
$f$'s one can construct the renormalization group
invariant constant   
by means of a finite multiplicative renormalization 
of $f$'s \cite {Kazakov}, \cite {GR}.}.  

Before we turn to the application of the QCD sum rule method let us first 
compare  the Operator Product Expansions (OPE) for  the quantity 
$B(Q^2)\equiv -\chi (Q^2)/Q^2$ in QCD 
and in YM theory. 
In  leading order only the  gluon fields contribute in both cases. 
The results of calculation
of these OPE's  can  be found in refs. 
\cite {NSVZ1}, \cite {KKP}, \cite {Narison}:
\begin{eqnarray}
B_{\rm YM}^{\rm OPE}(Q^2)=\left ({g^2\over
32\pi^2}\right )^2 Q^2 
\left ({2 \over \pi^2} \log{Q^2\over \mu^2_{\rm BLM}}+
{ D_4\over Q^4}+{D_6\over
Q^6}\right )+inst., \nonumber \\
B_{\rm QCD}^{\rm OPE}(Q^2)=\left ({g^2\over
32\pi^2}\right )^2 Q^2 
\left ({2\over \pi^2}\log {Q^2\over \mu^2_{\rm BLM}} 
+{ D'_4\over Q^4}+{D'_6\over
Q^6}\right )+inst., 
\end{eqnarray} 
where
\begin{eqnarray}
D_4=4 \langle 0|G^a_{\mu\nu}G^a_{\mu\nu}|0\rangle,~~~~D_6=8 g f^{abc}\langle
0|G^a_{\mu\alpha}G^b_{\alpha\beta}G^c_{\beta\mu}|0\rangle,
\nonumber \\
D'_4=D_4+O(\alpha_s m_q\langle {\bar q} q \rangle ),~~~
D'_6=D_6+O(\alpha^2_s \langle {\bar q} q \rangle ^2).
\end{eqnarray} 
The instanton contributions in eq. (6)  are suppressed as
$Q^{-n}$, where  $n \simeq 12$ \cite{NSVZ1}. Since for practical 
calculations we use eq. (3),
the NLO perturbative corrections are not explicitly shown  in eq. (6) 
for brevity.
Let us make some comments about  the quantity $B_{\rm QCD}^{\rm OPE}$.  
As we mentioned above, the perturbative part of this correlator in the
NLO is the same as that of $B_{\rm YM}^{\rm OPE}$.
However, there are nonperturbative contributions 
in $B_{\rm QCD}^{\rm OPE}(Q^2)$ which do not appear
in the expression for $B_{\rm YM}^{\rm OPE}(Q^2)$. Those are 
related to the quark condensate. The first such  contribution
modifies the $1/Q^4$ term in the OPE for $B_{\rm QCD}^{\rm OPE}$. 
The new contribution
is proportional to $\alpha_s m_q\langle {\bar q} q \rangle$.
The next nonperturbative 
correction related to the quark condensate appears at order
$1/Q^6$ and yields a new term in addition to the operator $O_6$.
The additional term is proportional to $\alpha_s^2 \langle {\bar q} q
\rangle ^2$. 

Now we can turn to the QCD sum rule analysis. It is useful to apply
the Borel transformation \cite {SVZ} 
(with the Borel parameter denoted by $M^2$) 
to eq. (2) with the phenomenological part
on its l.h.s. and the OPE on  its r.h.s.  
This leads to the following sum rules
\begin{eqnarray}
\int_0^{\infty}e^{-s/M^2} { \rho ^{poles}_{\rm YM}(s)\over s}ds=
\int_0^{s_{0}} e^{-s/M^2}{{\tilde  \rho}_{\rm YM}(s)\over
s}ds+ \pi \left ({g^2\over 32\pi^2}\right )^2 
\left (D_4+{D_6\over M^2}+...\right )\nonumber \\ +\chi_{\rm YM}(0), 
\nonumber \\
\int_0^{\infty}e^{-s/M^2} { \rho ^{poles}_{\rm QCD}(s)\over s}ds=
\int_0^{s_{1}} e^{-s/M^2}{ {\tilde\rho}_{\rm QCD}(s)\over
s}ds+ \pi \left ({g^2\over 32\pi^2}\right )^2     
\left (D'_4+{D'_6\over M^2}+...\right ). \nonumber
\end{eqnarray} 
Taking now the limit $M^2\rightarrow \infty$ we get\footnote{In this
way one derives the finite energy sum rules \cite {FESR} modified
by condensate contributions \cite {FESRM}.} 
\begin{eqnarray}
-\chi_{\rm YM}(0)+f^2_{G_0}m^2_{G_0}={a s_{0}^2\over 2}+ 
\pi \left ({g^2\over 32\pi^2}\right )^2 D_4, \\
f^2_{G}m^2_{G}+f^2_{\eta'}m^2_{\eta'}={a s_{1}^2\over 2}+ 
\pi \left ({g^2\over 32\pi^2}\right )^2 D'_4.
\end{eqnarray}

We have mentioned already that our goal is to study
how the QCD glueball mass differs from the one of quenched 
lattice gauge theory.
However,  eqs. (8,9)  alone  are  not yet
enough to determine whether that difference  really
exists. We need an additional relation.   
One way to get the new  relation is to
use the dispersion relation for the correlator of the 
topological susceptibility $\chi(Q^2)$  itself 
\begin{eqnarray}
\chi(Q^2)={1\over \pi}\int_0^{\infty}{\rho(s) ds
\over s+Q^2}+{\rm subtractions}. \nonumber
\end{eqnarray} 
In order to get rid of the subtractions let us use again the 
Borel transformation. Applying this transformation to
the correlator one gets the 
following relation 
\begin{eqnarray}
{1\over \pi}\int_0^{\infty}e^{-s/M^2} { \rho ^{poles}(s)}ds=
{1\over \pi}\int_0^{s_0}e^{-s/M^2}{ {\tilde \rho}(s)}ds-
\left ({g^2\over 32\pi^2}\right )^2 \left ({D_6}+O 
\left ( {1 \over M^2}\right ) \right ), \nonumber
\end{eqnarray} 
where we dropped  the subscripts distinguishing QCD from 
YM theory.
Taking the limit $M^2\rightarrow \infty$ we  obtain  the following relations 
\begin{eqnarray}
f^2_{G_0}m^4_{G_0}={a s_{0}^3\over 3}-\pi 
\left ({g^2\over 32\pi^2}\right )^2 D_6,~~~~~
f^2_{G}m^4_{G}+f^2_{\eta'}m^4_{\eta'}={a s_{1}^3\over 3}-\pi 
\left ({g^2\over 32\pi^2}\right )^2 D'_6.
\end{eqnarray} 

Having eqs. (8-10), one can use them 
to find the relations between the quantities
defined in  YM theory and in full QCD.
For that goal we are going to use the Witten-Veneziano 
arguments, but along with leading terms we keep corrections of order 
$1/N_c$.  In the large $N_c$ limit $f^2_{G}m^2_{G}\rightarrow
f^2_{G_0}m^2_{G_0}$,
and in accordance with the Witten-Veneziano (WV) relation \cite{Witten},
\cite{Venezia}\footnote{In our conventions  $\chi$ is
defined in Minkowski space-time and turns out to be a negative quantity. 
One can turn to Euclidean space-time in eq. (1) making the substitutions
$x_0\rightarrow -ix_4$ and $G\tilde G\rightarrow iG\tilde G$.
The Euclidean quantity, which is used in the lattice calculations, 
is a positive number  $\chi_{\rm YM}^{Eucl}(0)=-\chi_{\rm YM}^{Mink}(0)$.} 
$\chi_{\rm YM}(0)=-f^2_{\eta'}m^2_{\eta'}$.
As a result the left  hand sides of eqs. (8) and (9) are equal in the large
$N_c$ limit. On the other hand, the second terms on the r.h.s.
of eqs. (8,9) are also equal in that limit 
(difference between them is of order 
$1/N_c^2$). So, we come to the conclusion that
$\lim_{N_c\rightarrow \infty}(s_{1}-s_{0})=0$. 
The same result can be drawn from consideration of
eqs. (10) in the large $N_c$ limit. 

Since we are going to keep $1/N_c$ corrections as well as leading
order terms let us
introduce the following parametrization for the continuum thresholds
\begin{eqnarray}
s_{1}=s_{0}+{\delta\over N_c},
\end{eqnarray} 
where $\delta$ is an unknown quantity
which is of order of the unity in the large $N_c$ limit. 
In what follows we are going to solve the system 
of equations (8-11) keeping track of $1/N_c$ corrections. 
It is convenient to introduce the following notations
\begin{eqnarray}
{\tilde {D}_{4(6)}} \equiv \pi \left ({g^2\over 32\pi^2}\right )^2 
D_{4(6)},~~~ 
{\tilde{D}'_{4(6)}} \equiv \pi \left ({g^2\over 32\pi^2}\right )^2 
D'_{4(6)}.
\nonumber 
\end{eqnarray} 
Based on the rules of the large $N_c$ expansion one can see that
\begin{eqnarray}
{\tilde{D}'_{4}}-{\tilde{D}_{4}}\propto O 
\bigl ( {m_q},{1 \over N_c^2}\bigr ),~~~
{\tilde{D}'_{6}}-{\tilde{D}_{6}} \propto O 
\bigl ( {\langle {\bar q} q \rangle ^2 \over N_c^4}\bigr ).
\nonumber
\end{eqnarray} 
Using the notations  given above, and neglecting 
all contributions of order $1/N_c^2$ and higher, in the chiral limit 
one  can rewrite 
the system of eqs. (8-10) as follows
\begin{eqnarray}
-\chi_{\rm YM}(0)+f^2_{G_0}m^2_{G_0} ={a s_{0}^2\over 2}+ {\tilde {D}_4},~~~
f^2_{G_0}m^4_{G_0} ={a s_{0}^3\over 3}-{\tilde {D}_6}, \\
f^2_{G}m^2_{G}+f^2_{\eta'}m^2_{\eta'} =
{a s_{1}^2\over 2}+ {\tilde {D}_4},~~~ 
f^2_{G}m^4_{G}+
f^2_{\eta'}m^4_{\eta'} ={a s_{1}^3\over 3}-{\tilde {D}_6}.
\end{eqnarray} 
Before we  go  further let us make some comments.
 Two equations given in (12) contain
only two unknowns $f_{G_0}$ and $s_0$ (assuming that
the YM glueball mass $m_{G_0}$ is known from lattice calculations
\cite {UKQCD}). Hence, those two equations can in general be solved
and the values of $f_{G_0}$ and $s_0$ can uniquely be determined. 
This is done in the next section. 
On the other hand, the two equations in (13)
contain three unknowns, $f_{G}$ and $s_1$ and $m_G$. 
So one can not determine $m_G$ uniquely.
The only thing
one can do is to calculate the decay constant $f_{G}$
and mass $m_G$ for chosen values of the continuum threshold $s_1$
which are dictated by previous analyses of the flavor singlet
pseudoscalar channel. This calculation is also 
carried out below. Before we turn to the
numerical simulations we  can try to extract some analytic 
relations for the  QCD glueball mass and 
decay constant studying the system of equations (12-13).
The relation between the continuum threshold
parameters (11) allows one to set the equations (12-13) as following
\begin{eqnarray}
f^2_{G}m^2_{G}+f^2_{\eta'}m^2_{\eta'} =
-\chi_{\rm YM}(0)+f^2_{G_0}m^2_{G_0}
+{a \delta\over N_c}s_{0},\\ 
f^2_{G}m^4_{G}+
f^2_{\eta'}m^4_{\eta'} =f^2_{G_0}m^4_{G_0}+{a \delta\over
N_c}s^2_{0}.
\end{eqnarray} 
Now the key observation is that in pure YM theory there 
are not light meson states, thus  the decay
$G_0 \rightarrow 3\pi$ does not occur in this  theory.
On the other hand, this decay should easily go in QCD. 
Hence, the continuum threshold 
for QCD $s_1$ should be less
or equal to the continuum threshold for YM theory\footnote[5]{
I am grateful to Glennys Farrar for bringing this
line of arguments to my attention.} 
\begin{eqnarray}
s_1 \le s_0 ~~~{\rm or}~~~ \delta \le 0.
\end{eqnarray}
Thus, 
up to order  $1/N_c^2$ and in the chiral limit 
the following inequalities  exist between 
the quantities defined in pure YM theory and in full QCD 
\begin{eqnarray}
f^2_{G}m^2_{G}+f^2_{\eta'}m^2_{\eta'}\le 
-\chi_{\rm YM}(0)+f^2_{G_0}m^2_{G_0}, \\
f^2_{G}m^4_{G}+f^2_{\eta'}m^4_{\eta'}\le 
f^2_{G_0}m^4_{G_0}. 
\end{eqnarray} 
Let us point out that eq. (17)
yields the Witten-Veneziano relation  
in the limit of infinite $N_c$. Indeed, in the large $N_c$ limit 
$f^2_{G}m^2_{G}\rightarrow
f^2_{G_0}m^2_{G_0}$, $s_0 \rightarrow s_1$.   
Hence, in that  limit  eq. (17) is saturated
and  it turns into the Witten-Veneziano (WV) relation \cite{Witten},
\cite{Venezia} $\chi_{\rm YM}(0)= -f^2_{\eta'}m^2_{\eta'}$.
 
Let us rewrite  eq. (18) in the following form
\begin{eqnarray}
m_{G}^4 \le \left ( {f_{G_0}\over f_G} \right )^2 m_{G_0}^4-
\left ( {f_{\eta'}\over f_G} \right )^2 m_{\eta'}^4.
\end{eqnarray} 
This inequality allows one to calculate the upper bound on 
the pseudoscalar glueball  mass in QCD. 
The inequality shows that if the values for $f_{G_0}$ and $f_{G}$ 
are sufficiently close to each other, then  the QCD glueball 
is lighter than the glueball of YM theory. In the next section, 
based on numerical studies, we 
demonstrate that this indeed is the case. 

It is interesting to check the large $N_c$ behavior 
of eq. (19). Notice that,  since the anomaly 
disappears in the limit when  $N_c\rightarrow \infty$,  there is no 
flavor singlet meson -glueball mixing term anymore
in the  effective Lagrangian \cite {Trio} in that limit. 
One should  therefore  expect to have  equal masses for the 
glueballs in QCD and 
pure YM theory when $N_c\rightarrow \infty$. 
Recalling that $f_{\eta'}\propto \sqrt{N_c}$, $m_{\eta'}^2\propto 1/N_c$,
$f^2_{G_0}m_{G_0}^2\sim  1$ and 
substituting these into eq. (19) we get 
$$
m_{G_0}^2-m_{G}^2 \propto {1\over N_c},
$$
which is consistent with ones expectation.

\begin{center}
{\bf 3. Some Estimates and Predictions}
\end{center}

Let us now turn to  numerical estimates. First of all let us 
list  all the approximations we  made  deriving eqs. (12-15).
There are  scheme dependent  NLO 
perturbative corrections involved in the 
derivation. 
Besides that, we worked in the chiral limit
neglecting
$u,~d$ and, most importantly, $s$ quark masses. Hence, the  
$\eta-\eta'$ mixing 
and all other  nonsinglet pseudoscalar mesons are also 
neglected \footnote[1]{All these corrections are
expected to be of order  $O(m_{u,d,s}/\Lambda_{glueball})$,
where $\Lambda_{glueball}\simeq 1.5-2~{\rm GeV}$ is an  effective
scale above which  the existence of  glueballs become important for
hadron physics.}. Below  we show that the contributions of
the $\eta(1295)$, the  $\eta (1410)$ and the $\eta(1490)$ in 
the sum rules are rather small and can also be neglected. 

If one  knew all the numerical values 
for the quantities on  the r.h.s of eq. (19), 
one would be able to predict the upper bound on the 
QCD glueball mass. 
Unfortunately, $f_{G_0}$ and $f_{G}$
are not known. 
We can use however the sum rules derived in the previous section to
calculate the value for $f_{G_0}$. Indeed, consider the equations 
given in (12).
One can solve this system with
$f_{G_0}$ and $s_0$ treated  as unknowns. 
The numerical solution for 
that  system yields the following result
$f_{G_0}=(27\pm 3)~{\rm MeV}$ and $s_{0}=7.4
\pm 0.5~{\rm GeV^2}$.

In calculating the numbers
given above we have used the following numerical values  
for $f_{\eta'}=(29\pm 3)~{\rm MeV}$,
\footnote[2]{Which  corresponds to $F_0=(105\pm 12)~{\rm MeV} $ 
taken in  conventional normalization for the singlet axial current.} 
$m_{\eta'}=(957.77 \pm 0.14)~{\rm MeV} $,
$m_{G_0}=(2.3\pm 0.2)~{\rm GeV} $ \cite{UKQCD},
$\alpha_s(2~{\rm GeV})=0.33\pm 0.05 $. 
There are a number of estimates for the gluon condensate 
in the literature (see refs. \cite {SVZ}, \cite{gg}). We take
the world average value of these calculations 
$\langle{\alpha_s\over \pi} G_{\mu\nu}^2\rangle= (2.5\pm
0.9)~10^{-2}~{\rm GeV}^4$. The corresponding values
for $\tilde D_4$ and $\tilde D_6$ are 
$ {\tilde D_4}= (4.0\pm 1.7) 10^{-4}~{\rm GeV^4}$, 
${\tilde D_6}= (0.7 \pm 0.3 ) 10^{-4}~{\rm GeV^6}$.

There are also several  lattice estimates for the topological
susceptibility $\chi_{\rm YM}(0)$ (for a recent review see \cite{Shuryakrev}). 
For our estimates we take  the   
result of refs. \cite{Topsus}, $\chi_{\rm YM}^{Eucl}(0)=
(175\pm 5~{\rm MeV})^4$, 
which is also in good agreement with an earlier theoretical 
estimate \cite {Venezia}\footnote{In order to study whether our
results are sensitive to the numerical value of $\chi_{\rm YM}(0)$ 
we varied the value of this quantity  from $(100~{\rm MeV})^4$
to $(190~{\rm MeV})^4$. Outside of the interval given by
these numbers the  system of equations  does not have a solution. 
On the other hand, inside the interval the results are 
rather  insensitive  to the value of $\chi_{\rm YM}(0)$, varying 
$\chi_{\rm YM}(0)$,  
$s_0$ changes in the region $6.8 -8~{\rm GeV}$ and $f_{G_0}$
varies in the interval  $24-34~{\rm MeV}$.}.  

Now we can make the prediction for the 
matrix elements of the gluonic operator acting
on the pseudoscalar pure YM glueball state with mass 
$m_{G_0}=(2.3\pm 0.2)~{\rm GeV} $ 
\begin{eqnarray}
\langle 0|g^2  G^a_{\mu\nu}{\tilde G^a_{\mu\nu}}|G_0\rangle
=(45 \pm 9)~{\rm GeV}^3.
\end{eqnarray}
The uncertainty  in this  result  dominantly comes  from the error 
bars associated with the value for the topological susceptibility 
and also with the value of $f_{\eta'}$ \footnote[3]{The only available
lattice calculations for the matrix elements of the gluonic 
operators were presented in ref. \cite {Kentucky}. However, 
the predictions of ref. \cite {Kentucky} are in conflict
with the theoretical estimates in the case of $0^{++}$ channel (see
the discussion in ref.  \cite {Shuryak})  and also with our 
results for  $0^{-+}$ channel.}. The same matrix 
element for the $\eta'$ meson state has the following numerical value
$\langle 0|g^2  G^a_{\mu\nu}{\tilde G^a_{\mu\nu}}|\eta'\rangle
=(8.4 \pm 0.8 )~{\rm GeV}^3$.  The values of the matrix element
for the QCD glueball state $G$ will also be given below. 
Notice that
all the numerical results for the scale dependent quantities
such as the decay constants $f_{G_0}$, $f_{G}$ and the 
matrix element in eq. (20) are to be taken at 
the normalization point approximately equal to the value
of the glueball mass.  Also, all the results 
presented above should be given 10-20 percent 
systematic error bars 
associated with the approximations and the method we have used. 

As we mentioned already we  checked whether the presence 
of the $\eta(1295)$, the $\eta (1410)$ and the $\eta(1490)$ 
mesons could affect our results.
We have included the contributions of these resonances in the 
sum rules used above. In order to determine the 
decay constants  of these resonances 
we used the experimental data for  the production rate of these
states in $J/\psi$ radiative decays \cite{MarkIII48}, \cite{BESC}.
The values of the decay constants (defined as in eq. (5))
are rather small numbers:
$f_{\eta(1490)}= (7.5\pm 1.9)~{\rm MeV}$, $ f_{\eta(1410)}=
(4.8 - 6.7) ~{\rm MeV}$ and $f_{\eta(1295)}= (4.5\pm 1.0)~{\rm
MeV}$. Including these numbers into the sum rules 
one can see that the final results for the glueball 
mass and decay constants change unsubstantially. 

Now let us turn to the calculation of the 
mass and decay constant of the QCD pseudoscalar glueball.
Unfortunately, our method does not allow one to 
determine these values uniquely. The set of two equations 
given in (13) contains three unknowns, $f_G$, $m_G$ and $s_1$.
Below we present the results of numerical solutions of these
equations for different values of the continuum threshold
$s_1$. The values for $s_1$ are chosen between the upper bound
determined as  $s_1\le s_0=7.4~{\rm GeV}^2$ and the lower bound 
given by  $s_1\ge 3~ {\rm GeV}^2$
(below this value the continuum threshold comes very close to 
the resonance  mass square and applicability of the sum rule
approach breaks down).
The results are summarized in Table 1\footnote[8]{In  
Table 1 and also Table 2 below only the error bars associated with
the method of numerical calculations are given.}.

\begin{table}
  \begin{center}
    \begin{tabular}{ |l|r|r|r|r|r|}
    \hline
    $s_1~{\rm GeV^2}$ & $f_G~{\rm MeV}$ & $m_G~{\rm GeV}$ & 
     $s_1~{\rm GeV^2}$ & $f_G~{\rm MeV}$ & $m_G~{\rm GeV}$  \\ \hline
    7.4        & $ 29\pm 2$      &  $ 2.27\pm 0.04$ &         
    5.0        & $  21.5 \pm 2.5$    &  $ 1.9 \pm 0.05$      \\ \hline

    7.0        & $ 28 \pm 2$     &  $ 2.2\pm 0.04$  &
    4.5        & $  20 \pm 3$    &  $ 1.8 \pm 0.1$        \\ \hline

    6.5        & $ 26.5\pm 1.5$    &  $ 2.15\pm 0.05$  &      
    4.0        & $  18 \pm 3.5$    &  $ 1.73  \pm 0.12$     \\ \hline

    6.0        & $  25\pm 2$   &  $ 2.07 \pm 0.05$  &
    3.5        & $  16 \pm 4$    &  $ 1.61 \pm 0.14$        \\ \hline

    5.5        & $  23 \pm 2$  &  $ 1.97 \pm 0.05$  &
    3.0        & $  13.5 \pm 5.5$    &  $ 1.47  \pm 0.20$           \\ \hline
    
    \end{tabular}   
\caption {Sum rule results}    
\end{center}
\end{table}
As we see from the  table the QCD glueball is lighter than the 
glueball of YM theory. However,  it is hard
to identify the QCD glueball with the $\eta (1410)$.   
The very low value for the continuum threshold is needed
in order to have QCD  glueball mass at about 1.4 GeV.   

Though we can not  determine uniquely the value for the 
continuum threshold from our consideration, one can use 
the estimate given in ref. \cite {NSVZbig} $s_1\simeq 10 m_{\rho}^2
\simeq 6~{\rm GeV^2}$.
If this number  is accepted, then in accordance with  Table 1
the QCD glueball mass is $m_G= (2.07\pm 0.05\pm 0.3~(syst.))
~{\rm GeV}$ and decay constant $f_G= (25\pm 2\pm 4~(syst.)) ~{\rm MeV}$. 
 
Having these numbers at hand one can predict 
the $J/\psi$ decay width into a  glueball state and photon.
$J/\psi$ radiative decays are very effective tools 
in studying the spectroscopy on light mesons. In the present case we are
going to deal with the processes like $J/\psi\rightarrow
R(0^{-+})~\gamma $, where $R$ stands for the resonance being considered.
The theory of these decays was worked out in 
ref. \cite {NSVZJ}\footnote[4]{See also the  approach  
developed in \cite {CF}, \cite {CFL}.}.
One can consider  the ratio
\begin{eqnarray}
r\equiv {\Gamma(J/\psi\rightarrow G~\gamma)\over \Gamma(J/\psi\rightarrow 
\eta'~\gamma)}. \nonumber
\end{eqnarray}
This ratio is independent of  the $J/\psi$ meson wave function
and is completely  defined by the properties of the pseudoscalar
mesons produced in the decay.
Assuming that the decay dominantly goes through the exchange 
of the intermediate gluons in the pseudoscalar state, the ratio $r$ can
be rewritten as follows \cite {NSVZJ}
\begin{eqnarray}
r={|\langle 0|{g^2 } G^a_{\mu\nu}{\tilde G^a_{\mu\nu}}|G\rangle|^2 
\over |\langle 0|{g^2 } G^a_{\mu\nu}{\tilde G^a_{\mu\nu}}|
\eta'\rangle|^2}\Bigl({m_{J/\psi}^2-m_{G}^2 \over 
m_{J/\psi}^2-m_{\eta'}^2}\Bigr)^3+O(\alpha_s^2). \nonumber
\end{eqnarray}
Using the  numerical results listed above in the table  
one can calculate the ratio $r$, matrix element 
$k\equiv \langle 0|{g^2 } G^a_{\mu\nu}{\tilde G^a_{\mu\nu}}|G\rangle $
and decay width $\Gamma(J/\psi\rightarrow G~\gamma)$ for the 
QCD glueball state. These results are summarized in  Table 2. 

\begin{table}
  \begin{center}
    \begin{tabular}{ |l|r|r|r|r|}
    \hline
$f_G~{\rm MeV}$&$m_G~{\rm GeV}$&$k ~{\rm GeV^3}$&r ratio &$\Gamma ~{\rm
keV}$ \\ \hline
$29\pm 2$  &   $2.27\pm 0.04$ & $47.2\pm 5.0$ & $4.2\pm 1.1$  
& $1.57\pm 0.4$  \\ \hline
$28\pm 2$  &  $ 2.2\pm 0.04$  & $42.8\pm 4.7 $ & $4.2\pm 1.0$ 
&  $1.57\pm 0.37$   \\ \hline
$25 \pm 2$ &  $ 2.07\pm 0.05$ & $33.8\pm 4.5 $ & $3.7\pm 1.0$ 
&  $1.38\pm 0.37$  \\ \hline
$21.5\pm 2.5$   &  $ 1.9\pm 0.06$  & $24.5\pm 4.6$ & $2.8\pm 1.1$ 
&  $1.0\pm 0.4$  \\ \hline
$18\pm 3.5$   &  $ 1.73\pm 0.12$  & $17.0\pm 5.2 $ & $1.8\pm 1.2$ 
&  $ 0.67 \pm 0.45$   \\ \hline
$13.5\pm 5.5$   &  $ 1.47\pm 0.20$  & $9.2\pm 7.5 $ & $0.75\pm 1.4$  
&  $0.28  \pm 0.53$   \\ \hline
    \end{tabular}
\caption[]{Some predictions}    

\end{center}
\end{table}
The numerical values for the decay width of the 
$J/\Psi$ meson into the QCD glueball and photon 
are substantially large and this decay can be observed  
in recent experimental studies.
\begin{center}   
{\bf Conclusions} 
\end{center}

Let us summarize our results. We studied whether 
the pseudoscalar glueball mass in full QCD differs from 
the one determined in the quenched lattice calculations. 

An inequality which  sets the upper bound on the mass of 
the pseudoscalar glueball in QCD (eq. (19)) is derived. 
In order to calculate  that  bound numerically one needs to know 
the decay constant of the QCD $0^{-+}$ glueball and also 
the mass and  decay constant of the pure YM glueball. 

The decay constants  in this work are calculated
using  the QCD sum rule  approach,  
while the value for the YM glueball mass is  taken 
from lattice calculations. The value calculated for the 
decay constant of pure YM glueball is shown to be important
for selfconsistency checks of lattice results. 

We found numerically that the mass of the QCD glueball 
is less than the mass of the glueball of  pure  YM theory.
The values for the $0^{-+}$ QCD glueball mass and decay constant 
(for the phenomenologically preferred value of the continuum
threshold parameter) are: $m_G= (2.07\pm 0.05\pm 0.3~(syst.))
~{\rm GeV}$ and $f_G= (25\pm 2\pm 4~(syst.)) ~{\rm MeV}$.
If  these numbers are accepted, then there is no particle discovered
so far which might be identified with the QCD pseudoscalar glueball. 
Further experimental searches in the $2~{\rm GeV}$ region are needed. 
In this case the status of the $\eta(1410)$ is unclear. 

In order to help resolve this question,  we predict the production 
rate of the QCD glueball  in the radiative decay of the  $J/\psi$
meson, $\Gamma (J/\Psi\rightarrow G\gamma)$. 
For a $2.07~{\rm GeV}$ glueball $\Gamma (J/\Psi\rightarrow G\gamma)$
is about three or four times greater than for the $\eta'$ meson. 
Thus, the prediction  for the branching ratio for that  process 
is large enough to be studied experimentally. 
  
I am  grateful to Glennys R. Farrar for bringing this  
problem to my attention and for valuable  discussions and suggestions. 
The author wishes to thank H. Neuberger and T. DeGrand 
for useful discussions.
\begin{center}
{\bf Note added} 
\end{center}

After this work was done we become aware of the papers \cite
{NewNarison} where the QCD sum rule method was used to 
calculate glueball masses and decay constants 
in scalar, pseudoscalar and tensor
channels. The calculations in refs. \cite {NewNarison}  
are done (without referring to lattice 
results) using the optimization  procedure with respect to Borel 
parameters  and continuum thresholds.
Our results for the mass and decay constant
of the pseudoscalar glueball are in good agreement with 
predictions of \cite {NewNarison}. We are grateful
to M. Schwetz  for bringing refs. \cite {NewNarison} to our attention.

\end{document}